\documentclass[acmtog]{acmart}  
\acmSubmissionID{1165}

\acmJournal{TOG}

\copyrightyear{2025}
\acmYear{2025}
\setcopyright{acmlicensed}\acmConference[SIGGRAPH Conference Papers '25]{Special Interest Group on Computer Graphics and Interactive Techniques Conference Conference Papers }{August 10--14, 2025}{Vancouver, BC, Canada}
\acmBooktitle{Special Interest Group on Computer Graphics and Interactive Techniques Conference Conference Papers (SIGGRAPH Conference Papers '25), August 10--14, 2025, Vancouver, BC, Canada}
\acmDOI{10.1145/3721238.3730741}
\acmISBN{979-8-4007-1540-2/2025/08}

\usepackage[ruled]{algorithm2e} 
\usepackage{amsfonts}
\usepackage{amsmath}

\usepackage{array}
\usepackage{bbm}
\usepackage{booktabs}
\usepackage{color, colortbl}
\usepackage{cleveref}
\usepackage{enumitem}
\usepackage{float}
\usepackage[symbol]{footmisc}
\usepackage{mathtools}
\usepackage{mathrsfs}
\usepackage{multirow}
\usepackage{pifont}
\usepackage{setspace}
\usepackage{subcaption}

\usepackage{xfrac}

\SetAlFnt{\small}
\SetAlCapFnt{\small}
\SetAlCapNameFnt{\small}
\SetAlCapHSkip{0pt}

\DeclareUnicodeCharacter{3000}{  }
\DeclareUnicodeCharacter{2212}{-}

\newcolumntype{L}[1]{>{\raggedright\let\newline\\\arraybackslash\hspace{0pt}}m{#1}}
\newcolumntype{C}[1]{>{\centering\let\newline\\\arraybackslash\hspace{0pt}}m{#1}}
\newcolumntype{R}[1]{>{\raggedleft\let\newline\\\arraybackslash\hspace{0pt}}m{#1}}

\setlist[itemize]{noitemsep, topsep=0pt}
\setlist[enumerate]{noitemsep, topsep=0pt}

\newcommand{\ceil}[1]{\left\lceil #1 \right\rceil}

\newcommand{\parens}[1]{\left(#1\right)}
\newcommand{\braces}[1]{\left\{#1\right\}}
\newcommand{\bracks}[1]{\left[#1\right]}
\newcommand{\modulus}[1]{\left\vert#1\right\vert}
\newcommand{\norm}[1]{\left\Vert#1\right\Vert}

\definecolor{Gray}{gray}{0.9}

\newcommand{\method}{DuetGen}
\newcommand{\dataset}{DD100}
\newcommand{\datasetlong}{DD100 dataset}
\newcommand{\pA}{\mathcal{A}}
\newcommand{\pB}{\mathcal{B}}

\newcommand{\bulletitem}{\item[$\bullet$]}
\definecolor{modification_color}{rgb}{1.0, 0.0, 0.0}

\begin{document}
\title{\method: Music Driven Two-Person Dance Generation via Hierarchical Masked Modeling}

\author{Anindita Ghosh*}
\thanks{*Work done during internship at Snap Inc.}

\affiliation{%
 \institution{DFKI, MPI for Informatics, SIC}
 \country{Germany}
 }
\email{anindita.ghosh@dfki.de}
\author{Bing Zhou}
\affiliation{%
\institution{Snap Inc.}
\country{USA}
}
\email{ bzhou@snapchat.com}
\author{Rishabh Dabral}
\affiliation{%
 \institution{MPI for Informatics, SIC}
 \country{Germany}
}
\email{rdabral@mpi-inf.mpg.de}
\author{Jian Wang}
\affiliation{%
 \institution{Snap Inc.}
 \country{USA}}
\email{jwang4@snapchat.com}
\author{Vladislav Golyanik}
\affiliation{%
 \institution{MPI for Informatics, SIC}
 \country{Germany}
}
\email{golyanik@mpi-inf.mpg.de}
\author{Christian Theobalt}
\affiliation{%
 \institution{MPI for Informatics, SIC}
 \country{Germany}}

\email{theobalt@mpi-inf.mpg.de}
\author{Philipp Slusallek}
\affiliation{%
 \institution{DFKI, SIC}
 \country{Germany}}

\email{philipp.slusallek@dfki.de}
\author{Chuan Guo}
\affiliation{%
 \institution{Snap Inc.}
 \country{USA}}
\email{cguo2@snapchat.com}

\renewcommand\shortauthors{Ghosh, A. et al}

\begin{abstract}

We present \method, a novel framework for generating interactive two-person dances from music. The key challenge of this task lies in the inherent complexities of two-person dance interactions, where the partners need to synchronize both with each other and with the music. Inspired by the recent advances in motion synthesis, we propose a two-stage solution: encoding two-person motions into discrete tokens and then generating these tokens from music.
To effectively capture intricate interactions, we represent both dancers' motions as a unified whole to learn the necessary motion tokens, and adopt a \textit{coarse-to-fine} learning strategy in both the stages. Our first stage utilizes a VQ-VAE that hierarchically separates high-level semantic features at a coarse temporal resolution from low-level details at a finer resolution, producing two discrete token sequences at different abstraction levels. Subsequently, in the second stage, two generative masked transformers learn to map music signals to these dance tokens: the first producing high-level semantic tokens, and the second, conditioned on music and these semantic tokens, producing the low-level tokens. We train both transformers to learn to predict randomly masked tokens within the sequence, enabling them to iteratively generate motion tokens by filling an empty token sequence during inference.
Through the hierarchical masked modeling and dedicated interaction representation, \method~ achieves the generation of synchronized and interactive two-person dances across various genres.
Extensive experiments and user studies on a benchmark duet dance dataset demonstrate state-of-the-art performance of \method~ in motion realism, music-dance alignment, and partner coordination.
Code and model weights are available at \href{https://github.com/anindita127/DuetGen}{https://github.com/anindita127/DuetGen}.
\end{abstract}

%
%


%
%


\begin{teaserfigure}
    \centering
    \includegraphics[width=\columnwidth]{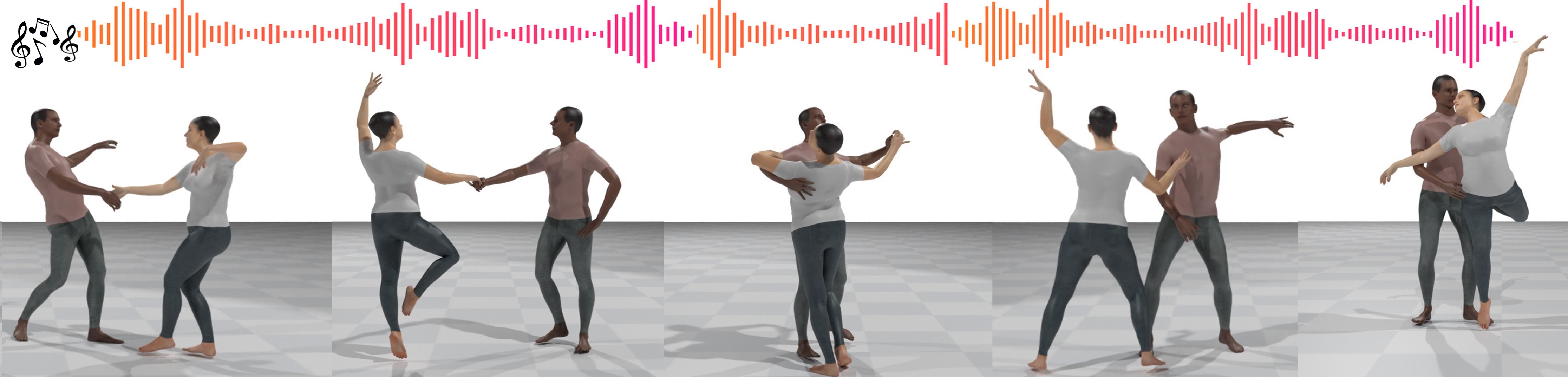}
    \caption{\textbf{\method} generates synchronized two-person dance choreography from input music, featuring natural and close interactions between dancers.  
    }
    \label{fig:teaser}
\end{teaserfigure}
\maketitle
\section{Introduction} \label{sec:introduction}
Duet dancing is a fundamental aspect of human culture that embodies coordination and mutual expression between partners. As one of the most interactive and visually engaging dance forms~\cite{10.1145/3323335}, it holds a significant presence across traditional and contemporary art.
With the proliferation of contemporary art in digital platforms today, the capability to synthesize dances, including duet dances, fosters new opportunities for creative content generation in domains spanning entertainment and communication media, virtual social interactions, and virtual education, among others.
While recent advances in text-based motion generation~\cite{ghosh2021text, petrovich2022temos, guo2024momask} and editing~\cite{li2025simmotionedit, athanasiou2024motionfix} allow users to choreograph dance through natural language, music offers a more natural and temporally aligned modality for generating expressive and synchronized duet performances~\cite{gong2023tm2d}.

Unlike single-person dances~\cite{tseng2022edge, alexanderson2023listen, bhattacharya2024danceanyway}, where timing and spatial consistency concern only one body, two-person dance synthesis gives rise to several key challenges in motion modeling. It requires maintaining synchronized motions between the dancers to ensure cohesive rhythm, tempo, and mutual positioning (as illustrated in~\Cref{fig:teaser}).
It further demands intricate modeling of lead-follow dynamics and close interactions, including twirls, lifts, and partner supports across diverse dance styles --- all while aligning both dancers' motions with the music.
Recent works have explored group-dance synthesis~\cite{le2023music, le2023controllable, yao2023dance}, 
focusing on synchronized multi-person dance motions. However, these methods do not address close interactions essential in duet dancing. While Duolando~\cite{siyao2023duolando} pioneered dance generation in two-person scenarios, it is limited to reactive partner motions rather than simultaneously generating motions of both dancers.

Towards this goal, we introduce \textit{\method}, the \textit{first} method designed to generate synchronized and interactive two-person dance motions from music.
Our key insights are twofold. First, we propose a unified representation for two-person motions that treats both dancers' movements as a whole, rather than modeling individual motions separately using single-person representations~\cite{javed2024intermask,siyao2023duolando}.
This unified representation proves especially effective in reducing the modeling complexity of inter-person interactions. Second, we propose a hierarchical learning of the unified dance motions, encoding movements into two levels of discrete motion tokens representing global semantics and fine details, respectively. This allows for efficient learning of the individual and interactive motions, and synchronizing them with music.
To implement our approach, we train a hierarchical VQ-VAE model to learn motion tokens at the two levels. We then employ two transformers to generate the motion tokens from music. The first transformer generates high-level semantic tokens from music, and the second transformer, conditioned on both music and the high-level tokens, generates the low-level tokens. We train both transformers using a masked modeling approach such that during inference, the transformers can iteratively predict a complete sequence of motion tokens from a fully-masked masked sequence.
We also incorporate a root motion predictor to refine the global root trajectories in the generated motions.

In summary, our main contributions are the following.
\begin{itemize}
    \item \method, the first framework for generating interactive two-person dance motions directly from music.
    \item A technique to model two-person dance motions in the discrete space with a hierarchy of abstractions, using our unified two-person representation, multi-scale motion quantization and two-stage generative masked transformers.
    \item Extensive experiments and user studies on the benchmark \datasetlong~\cite{siyao2023duolando}, demonstrating our effectiveness in synthesizing realistic two-person dances.
\end{itemize}

\begin{figure*}[t]
    \includegraphics[width=\textwidth]{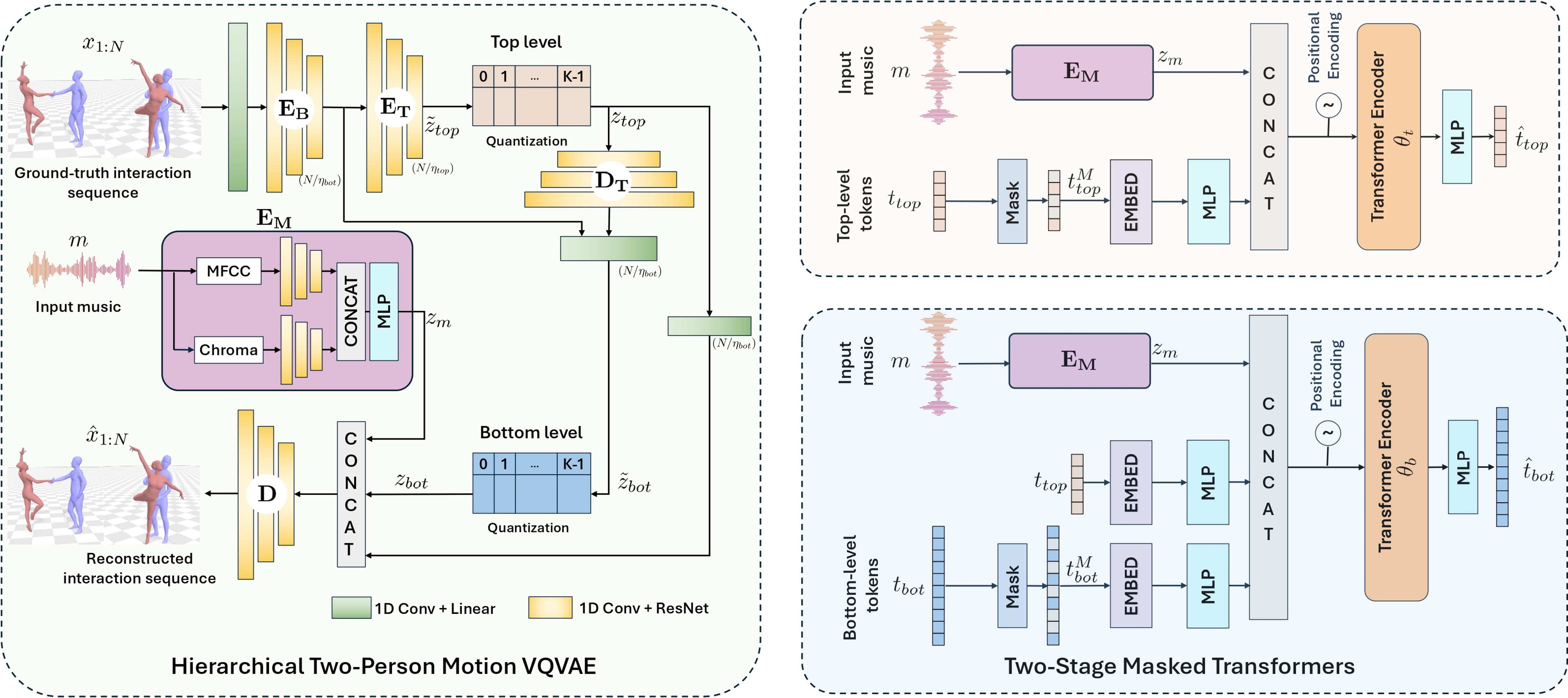}
    \caption{\textbf{\method~ Training Framework.} \textit{Left:} Our hierarchical two-person motion VQ-VAE encodes a unified two-person motion sequence $x$ of length $N$ into two-scale discrete token sequences. Top-level tokens at a coarse temporal resolution ($N/\eta_{top}$) capture global semantics, and bottom-level tokens at finer temporal resolution ($N/\eta_{bot}$) capture complementary low-level details.  
    \textit{Right:} We employ two transformers to model tokens of different granularity via masked modeling. The first-stage transformer (\textit{top-right}) predicts the masked top-level tokens from music. Subsequently, the second-stage transformer (\textit{bottom-right}) predicts the masked bottom-level tokens conditioned on both the music and the top-level token sequence.
    \label{fig:method}
    }
\end{figure*}

\section{Related Work} \label{sec:related_work}
We briefly summarize prior work in the related areas of motion quantization, multi-person motions, and music-to-dance synthesis.

\paragraph{Deep Motion Quantization}
Aristidou et al. ~\cite{aristidou2018deep} utilize triplet contrastive learning to convert continuous motions into semantically meaningful discrete motif words. Vector-quantized VAE (VQ-VAE)~\cite{van2017neural} was first applied to human motion modeling in TM2T~\cite{guo2022tm2t}, where autoencoded motion latent codes were quantized to indexed entries in a learnable codebook. T2M-GPT~\cite{zhang2023generating} enhanced this approach through exponential moving average and code reset techniques, which was then adopted by subsequent works~\cite{jiang2023motiongpt,zhang2024motiongpt}. 
SynNsync~\cite{maluleke2024synergy} quantizes full-body motion by disentangling motion into its
constituent pose, orientation, and translation via a parametric body model.
To reduce quantization error, MoMask~\cite{guo2024momask} implemented residual quantization (RVQ)~\cite{borsos2023audiolm} and used multiple layers to iteratively quantize a vector and its residuals. While RVQ achieves improved motion-space mapping, its quantization layers operate at full temporal scale and the tokens describe quantization residuals that have no specific motion semantics. In contrast, we follow a coarse-to-fine strategy that naturally separates high-level semantic features from low-level details, and represent them as motion tokens at different temporal resolutions.

\paragraph{Multi-Person Motion Synthesis}
Synthesizing motions for multiple persons presents unique challenges in modeling their interactions~\cite{sui2025survey}.
Starke et al.~\cite{starke2020local, starke2021neural} pioneered the use of mixture-of-expert networks to autoregressively predict poses based on inter-character features and instant control signals. Guo et al.~\cite{guo2022multi} introduced transformer-based cross-attention networks to forecast intensive interactions, such as combat, from historical observations.
A separate body of research has emerged on reactive motion synthesis~\cite{chopin2023interformer,ghosh2024remos,xu2024regennet,liu2023interactive,siyao2023duolando}, where one character's movements are generated in response to others. Notable contributions include InterFormer's skeleton-aware spatial attention~\cite{chopin2023interformer} and the integration of human-object interactions~\cite{liu2023interactive}.
More recently, ReMoS~\cite{ghosh2024remos} and ReGenNet~\cite{xu2024regennet} have improved the generation of interpersonal dynamics using combined diffusion-transformer architectures and interaction losses.
Text-driven multi-person motion generation has also gained traction, evolving from approaches that fine-tune single-person diffusion priors~\cite{shafir2023ComMDM,wang2024intercontrol} to more sophisticated methods~\cite{liang2024intergen,ponce2024in2in} enabled by large datasets like InterHuman~\cite{liang2024intergen} and Inter-X~\cite{xu2024inter}. Recent works~\cite{shan2024opendomain,fan2024freemotion} have extended these capabilities to more than two persons through specialized transformer architectures and pose-based intermediaries.
An emerging alternative leverages discrete representations, typically in a two-stage process: motion-to-token conversion~\cite{van2017neural} followed by token generation. This approach has shown promise in applications such as Duolando's audio-guided partner dance generation~\cite{siyao2023duolando} and InterMask's text-driven two-person motion synthesis~\cite{javed2024intermask}. We further extend this approach to duet dances using a unified representation of two-person motions and a hierarchical tokenization.

\paragraph{Music-Driven Dance Synthesis}
Traditional music-to-dance synthesis approaches~\cite{arikan2002interactive,au2022choreograph,chen2021choreomaster,shiratori2006dancing} relied on motion graphs and dynamic time warping to retrieve matching motion clips from existing libraries. While these methods generate plausible movements, they are constrained by their reliance on pre-existing motions and the need to maintain extensive motion libraries, limiting their scalability. The advent of deep learning shifted the paradigm toward data-driven approaches, framing music-to-dance as an autoregressive problem and solving it using RNNs and transformers~\cite{li2020learning,sun2022you,li2021ai,huang2020dance}. While these models effectively learned movement patterns and music correspondences, their deterministic nature often led to repetitive or frozen motions. Subsequent explorations on deep generative models, particularly with GAN-based approaches~\cite{li2022danceformer,sun2020deepdance} and hybrid GAN-VAE models~\cite{Dancing_2_music} for composing long-term dances from shorter clips, alleviated some of these limitations.
A significant breakthrough came with diffusion models~\cite{ho2020denoising} and GPT models~\cite{brown2020language}, which substantially improved motion quality and realism~\cite{siyao2022bailando,alexanderson2023listen,tseng2022edge}. This success led to specialized architectures like directional autoregressive diffusion~\cite{zhang2024bidirectional}, coarse-to-fine generation~\cite{li2024lodge}, and full-body synthesis~\cite{li2023finedance}. Bailando~\cite{siyao2022bailando} mapped dance movements to discrete tokens using VQ-VAE and generated them autoregressively with GPT. This discrete representation framework was extended by TM2D~\cite{gong2023tm2d} to achieve music-to-dance synthesis with further controllability from text prompts by sharing the codebooks. MoFusion~\cite{dabral2022mofusion}, and UDE~\cite{zhou2023ude} also approached this bimodality driven 3D dance generation, but with diffusion models. Nevertheless, these works remain limited to single-person dance synthesis.
While several works have explored generating group choreography from music~\cite{le2023controllable,wang2022groupdancer,le2023music}, they typically neglect close interactions between dancers. Duolando~\cite{siyao2023duolando} and InterDance~\cite{li2024interdance} made initial progress by generating follower movements in response to a given leader's sequence, 
with InterDance having the capability to generate two-person motion from audio by replacing the leader's motion with random noise.
Our work advances this field by simultaneously generating coordinated movements and close interactions for both dancers attuned to music signals.

\section{Method} \label{sec:method}

To generate two-person dance motions from music signal, we first model the dance motions in a discrete space. ~\Cref{fig:method} illustrates the training pipeline of our method, which consists of two main components: a hierarchical motion VQ-VAE (\Cref{sec:method_quantization}) that transforms motions into discrete tokens, and two-stage masked transformers (\Cref{sec:method_transformer}) that maps input music to these tokens. After training, the full music-to-dance mapping is achieved through the inference procedure shown in ~\Cref{fig:inference}.

\subsection{Data Representation}
\label{sec:method_rep}

\paragraph{Unified Two-Person Motion Representation}
Without loss of generality, we denote the two partners in a duet dance as $\pA$ and $\pB$ (note that the designations $\pA$ and $\pB$ are interchangeable). We represent the motions of a duet dance as $x_{1:N} \in \mathbb{R}^{N \times D}$ with $N$ frames and $D$ dimensional motion features that combine the motions of $\pA$ and $\pB$. 
We adopt the SMPL~\cite{loper2023smpl} body structure with $J=22$ body joints.
The unified two-person interaction representation for the $i^{th}$ motion frame consists of
\begin{equation}
    x_i = \bracks{
        t^{\delta}_\pA, r^{g}_\pA, j^{p}_\pA, j^{r}_\pA, j^{v}_\pA, c^{f}_\pA,
        t^{\epsilon}_\pB , r^{g}_\pB, j^{p}_\pB, j^{r}_\pB,j^{v}_\pB, c^{f}_\pB
    }_i,
\end{equation}
where $t^{\delta}_\pA \in \mathbb{R}^3$ is the root translation difference for $\pA$ from \textit{their own previous frame}, $t^{\epsilon}_\pB\in \mathbb{R}^3$ is the root translation difference for $\pB$ \textit{relative to $\pA$'s current frame}, $r^{g}_\pA, r^{g}_\pB \in \mathbb{R}^6$ are the global orientations of the root joint $\pA$ and $\pB$,
$j^{p}_\pA, j^{p}_\pB \in \mathbb{R}^{(J-1) \times 3}$ are their root-invariant local joint positions, $j^{r}_\pA, j^{r}_\pB \in \mathbb{R}^{(J-1)\times 6}$ are their 6D local joint rotations, $j^{v}_\pA, j^{v}_\pB \in \mathbb{R}^{J \times 3}$ are their local joint velocities, and $c^{f}_\pA, c^{f}_\pB \in \mathbb{R}^{4}$ are their binary foot-ground contact indicators obtained by thresholding the heel and toe joint velocities. Together, these features add up to the motion feature dimension $D = 536$.

Note that we deliberately represent $\pA$'s global position in the global coordinates as velocity and represent $\pB$'s global position relative to $\pA$'s. This \textit{relational global positioning} ensures that $\pB$'s global position is informed by their partner, making its prediction more manageable for networks. Our experiments demonstrate the key role of this approach in both VQ learning (\Cref{tab:reconstructions}) and the final generation (\Cref{tab:quant_baseline}). 

\paragraph{Music Features} 

We sample the input music $m$ at a rate that matches the temporal resolution of the motion sequence. We then extract MFCC~\cite{hossan2010novel}, MFCC delta, and Chroma features~\cite{shah2019Chroma} using the \textit{Librosa} library. The MFCC features are 40-dimensional and the Chroma features are 12-dimensional.

\subsection{Hierarchical Two-Person Motion Quantization}
\label{sec:method_quantization}
As illustrated in~\Cref{fig:method} (left), we discretize two-person dance motions into a hierarchy of tokens. The top-level tokens model high-level motion semantics, \textit{e.g.}, actions such as walking and turning induced by overall body movements, and bottom-level tokens represent fine-grained details, \textit{e.g.}, intricate articulations specific to a dance choreography. To obtain these tokens, our VQ-VAE encoder network consists of a bottom encoder $\mathbf{E_B}$ followed by a top encoder $\mathbf{E_T}$. Both $\mathbf{E_B}$ and $\mathbf{E_T}$ consist of 1D convolution blocks and they temporally downscale the input motion $x_{1:N}$ by factors of $\eta_{bot}$ and $\eta_{top}$ respectively, where $\eta_{top} > \eta_{bot} > 1$. $\mathbf{E_T}$ yields latent code sequences $\tilde{z}_{top} \in \mathbb{R}^{(N/\eta_{top}) \times D_{top}}$, where $D_{top}$ denotes the spatial dimensionality. We quantize this code sequence by replacing each of the code with its nearest entry in a learnable codebook $\mathcal{C}_{top}=\braces{c^t_k}_{k=1}^{K} \subset \mathbb{R}^{D_{top}}$ consisting of $K$ codes, resulting in the quantized vector sequence $z_{top} \in \mathbb{R}^{(N/\eta_{top}) \times D_{top}}$. Specifically, 
\begin{equation}
       z_{top} = \mathbf{Q_T}(\tilde{z}_{top}) ; \hspace{4mm} \tilde{z}_{top} = \mathbf{E_T}(\mathbf{E_B}(x_{1:N})),
\end{equation}
where $\mathbf{Q_T(\cdot)}$ denotes the quantization operation at the top level.

Subsequently, we upscale the top-level quantized features $z_{top} \in \mathbb{R}^{(N/\eta_{top}) \times D_{top}}$ by a factor of $\eta_{top} / \eta_{bot}$ through the top decoder $\mathbf{D_T}$. We combine them with the output features from $\mathbf{E_B}$ to produce latent features at a finer temporal resolution $\tilde{z}_{bot} \in \mathbb{R}^{(N/\eta_{bot}) \times D_{bot}}$, where $D_{bot}$ denotes the spatial dimensionality. We quantize these bottom-level latent codes using a separate codebook $\mathcal{C}_{bot}=\braces{c^b_k}_{k=1}^{K} \subset \mathbb{R}^{D_{bot}}$ where $K$ is the number of codes in the codebook, yielding quantized feature sequences $z_{bot} \in \mathbb{R}^{(N/\eta_{bot}) \times D_{bot}}$, as
 \begin{equation}
       \tilde{z}_{bot} = \parens{\mathbf{E_B}\parens{x_{1:N}}, \mathbf{D_T}\parens{z_{top}}} ; \hspace{4mm}
    z_{bot} = \mathbf{Q_B}(\tilde{z}_{bot}),
\end{equation}
where $\mathbf{Q_B}(\cdot)$ indicates the bottom-level quantization process.

We then concatenate the quantized features $z_{top}$ and $z_{bot}$, and decode them back to the motion sequence $\hat{x}_{1:N} \in \mathbb{R}^{N \times D}$ via the decoder $\mathbf{D}$. 
Interestingly, we observe that conditioning the decoder $\mathbf{D}$ additionally on the music features $z_m$ encourages faster convergence of the training and better VQ-VAE reconstructions (\Cref{tab:reconstructions}). Overall, the final reconstructed two-person sequence is given as

 \begin{equation}
     \hat{x}_{1:N} = \mathbf{D}\parens{z_{top}, z_{bot}, z_m}.
 \end{equation}

\begin{figure*}[t]
    \includegraphics[ width=\textwidth]{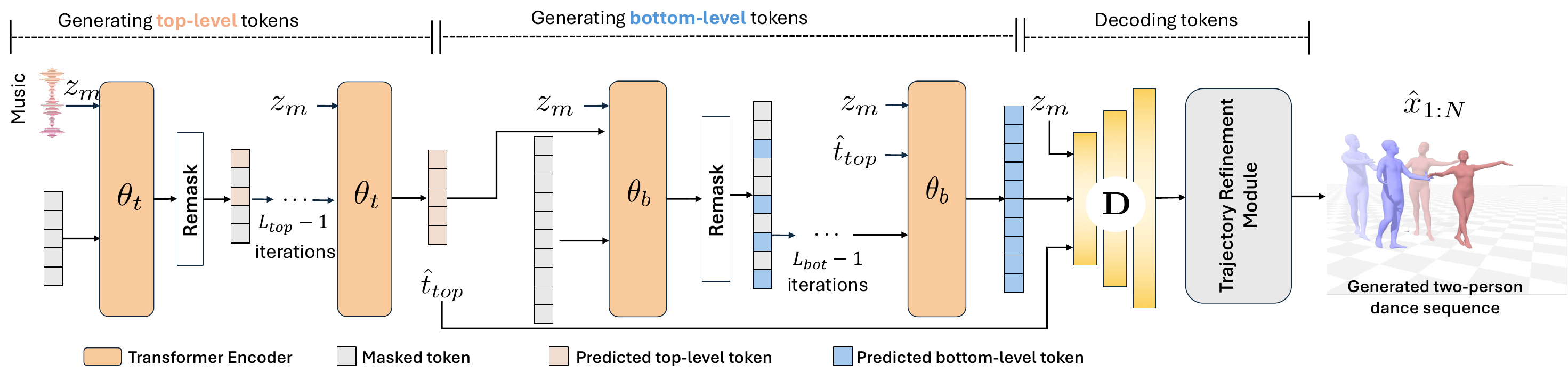}
    \caption{\textbf{Inference Process.} Our first-stage transformer $\theta_t$ iteratively fills an empty sequence of top-level tokens in $L_{top}$ iterations based on input music. Then, the second-stage transformer $\theta_b$ generates the complete sequence of bottom-level tokens in $L_{bot}$ iterations, conditioned on both music and the generated top-level tokens $\hat{t}_{top}$. We combine these token sequences and decode them into two-person dance motions via the VQ-VAE decoder $D$. We then apply a lightweight network to refine the root trajectories and produce the final outputs. For brevity, we only show the main network modules here. }
    \label{fig:inference}
\end{figure*}

\paragraph{Training Objectives}
We train the hierarchical VQ-VAE framework end-to-end. Our primary VQ-VAE training objective consists of a motion reconstruction loss $\mathcal{L}_r=\norm{x - \hat{x}}_2$ and a motion velocity loss $\mathcal{L}_v=\norm{\Delta{x} - \Delta{\hat{x}}}_2$. Following T2M-GPT\cite{zhang2023generating}, we use the exponential moving average (EMA) and codebook reset (Code Reset) techniques for both $\mathcal{C}_{top}$ and $\mathcal{C}_{bot}$ codebooks. We optimize for commitment losses on the two quantized latent sequences, as
\begin{equation}
    \mathcal{L}_{com} = 
    \beta_1 \norm{\tilde{z}_{top} - sg\bracks{z_{top}}}_2 + \beta_2 \norm{\tilde{z}_{bot} - sg\bracks{z_{bot}}}_2,
    \label{eqn:com_loss}
\end{equation}
where $\beta_1$ and $\beta_2$ are hyper-parameters for the commitment loss  and $sg[\cdot]$ is the stop-gradient operator.

While optimizing for overall character movements, the aforementioned losses do not enforce synchronization fidelity between the two dancers. Therefore, we apply feature transformation and forward kinematics $fk(\cdot)$ to the ground-truth motions features $x$ and the reconstructed motion features $\hat{x}$ to obtain the joint positions of $\pA$ and $\pB$ in the global coordinates, as $\parens{p_\pA, p_\pB}=fk(x)$ and $\parens{\hat{p}_\pA, \hat{p}_\pB}=fk(\hat{x})$ respectively. We then enforce reconstruction loss on the global-coordinate joint positions and relative joint distances between the two-persons, as
\begin{align}
    \mathcal{L}_{fk} &= \norm{p_\pA - \hat{p}_\pA}_2 + \norm{p_\pB - \hat{p}_\pB}_2, \\
    \mathcal{L}_{rel} &= \frac{1}{J} \sum_{j \in J} \lambda_j \sum_{k \in J} e^{-d\parens{p_{\pA_j}, p_{\pB_k}}} \modulus{d\parens{p_{\pA_j}, p_{\pB_k}} - d\parens{\hat{p}_{\pA_j}, \hat{p}_{\pB_k}}},
    \label{eqn:rel_loss}
\end{align}
where $d(\cdot,\cdot)$ is the Euclidean distance computed between each joint positions of $\pA$ and $\pB$, and $\lambda_j$ is a variable weight applied to individual joints with higher values for the end-effectors to better learn their rapid movements. We also use an exponentially decaying distance-aware weight $e^{-d\parens{p_{\pA_j}, p_{\pB_k}}}$ at each frame to emphasize joints that are closer to the other person and therefore more relevant for interactions~\cite{ghosh2024remos}.
    
Overall, we train our hierarchical two-person VQ-VAE to minimize the weighted sum of all loss terms, as
    \begin{equation}
        \mathcal{L}_{vq} = \lambda_{r}\mathcal{L}_{r} + \lambda_{v}\mathcal{L}_{v} + 
        \lambda_{com}\mathcal{L}_{com} + 
        \lambda_{fk}\mathcal{L}_{fk} +
        \lambda_{rel}\mathcal{L}_{rel},
        \label{eqn:total_loss}
    \end{equation}
where $\lambda_{r}$, $\lambda_{v}$, $\lambda_{com}$, $\lambda_{fk}$ and $\lambda_{rel}$ are scalar weights to balance the individual loss components.

\paragraph{Discrete Representation.} After training, we can represent each code in the quantized sequence at the top level, $z_{top}$, by its corresponding index $k$ in the codebook $\mathcal{C}_{top}$, thus transforming a two-person motion sequence into a discrete token sequence $t_{top}\in\braces{0, 1, ...,K}^{N/\eta_{top}}$. Similarly, we obtain the bottom-level tokens as $t_{bot} \in \braces{0, 1, ..., K}^{N/\eta_{bot}}$. We also retain the ability to decode these token sequences back to two-person motions.
Our following two-stage transformers learn to generate this hierarchy of discrete tokens from music through generative masked modeling.

\subsection{Music to Motion Via Two-Stage Masked Transformers}
\label{sec:method_transformer}
We use two-stage bidirectional transformers to map input music signals $m$ to the discrete two-person dance token sequences $t_{top}$ and $t_{bot}$ (\Cref{fig:method}, right). To this end, we follow a generative masked modeling approach. First, we randomly mask out a varying fraction of sequence elements by replacing the original tokens with a special $\bracks{\mathrm{MASK}}$ token. We then train the transformers to predict these masked tokens given the token context and relevant conditions. Following MaskGIT~\cite{chang2022maskgit}, we adopt a cosine function for scheduling the masking ratio as $\gamma(\tau) = \cos \parens{\frac{\pi \tau}{2}} \in \bracks{0,1}$ where $\tau \sim \mathcal{U}\parens{0,1}$.
During training, we randomly sample $\tau$ at each iteration, and uniformly select $\ceil{\gamma(\tau) \cdot N}$ tokens from the token sequence of length $N$ for masking.
To further enhance the contextual reasoning of the masked transformer, we adopt the re-masking strategy~\cite{guo2024momask} used in BERT pre-training~\cite{devlin2018bert}, where we either replace a small portion of masked tokens with random tokens or restore them to their original values.

\paragraph{Modeling Top-Level Motion Tokens}
We first encode the music signal through the music encoder $\mathbf{E_M}$ to obtain music features $z_m$.
Following the masking strategy, we mask out portions of the top-level tokens to get $t^M_{top}$, which we embed and then concatenate with $z_m$. After positional encoding, we provide these features into a transformer encoder $\theta_t$, which predicts the discrete distribution of tokens at the masked positions. Our objective is to minimize the negative log-likelihood of the prediction on masked tokens, as
\begin{equation}
    \mathcal{L}_{tmask} = \sum_{{t^M_{top}}_k=\bracks{\mathrm{MASK}}} -\log \theta_{t} \parens{{t_{top}}_k \Big| t^M_{top}, m}.
\end{equation}

\paragraph{Modeling Bottom-Level Motion Tokens}
The bottom-level tokens are modeled by a separate transformer $\theta_b$ in a similar fashion, with the addition of conditioned on the associated top-level tokens.
Following the same masking strategy, we mask portions of the bottom-level tokens to get $t^M_{bot}$. The training objective is to minimize the negative log-likelihood of the predicted masked tokens conditioned on the bottom-level token contexts $t^M_{bot}$, full sequence of top-level tokens $t_{top}$, and music signals $m$:
\begin{equation}
    \mathcal{L}_{bmask} = \sum_{{t^M_{bot}}_k=\bracks{\mathrm{MASK}}} -\log \theta_{b} \parens{{t_{bot}}_k \Big| t^M_{bot}, m, t_{top}}.
\end{equation}

For both transformers, we adopt a classifier-free guidance strategy (CFG)~\cite{chang2022maskgit}, where we train the transformers without music feature conditioning for $10\%$ of the time.

\subsection{Inference} \label{inference}
During inference, we generate a two-person dance sequence $\hat{x}$ of length $N$ from input music (\Cref{fig:inference}). 
We start from an empty sequence of length $(N/\eta_{top})$, where all the tokens are masked, and use $\theta_t$ conditioned on music feature $z_m$ to generate the top-level token sequence in $L_{top}$ iterations.
At each iteration $l$, $\theta_t$ predicts the probability distribution of tokens at the masked locations and samples motion tokens accordingly.
We re-mask the sampled tokens with the lowest $\ceil{\gamma\parens{\frac{l}{L_{top}}}\cdot\parens{\frac{N}{\eta_{top}}}}$ probabilities, while keeping the other tokens unchanged for subsequent iterations. This process repeats until iteration $L_{top}$, yielding the complete sequence of top-level tokens $\hat{t}_{top}$. Next, we use  $\theta_b$ to generate the bottom-level token sequence $\hat{t}_{bot}$ of length $(N/\eta_{bot})$ through $L_{bot}$ iterations similarly, conditioned on both $z_m$ and $\hat{t}_{top}$. Throughout this process, we apply classifier-free guidance at the projection layer of the transformers with guidance scale $s$. Finally, we decode these two motion token sequences to the motion space using the VQ-VAE decoder $\mathbf{D}$.

\paragraph{Trajectory Refinement Module}
While our method generates plausible dance motions, the results exhibit noticeable sliding issues, primarily due to inaccurate root motions. Since global root motions can generally be determined from local body movements~\cite{guo2024generative}, we implement a regressor with 1D convolutional layers.
This regressor takes root orientations and local motion features
$
    x^{local}_i = \bracks{ r^{g}_\pA, j^{p}_\pA, j^{r}_\pA, j^{v}_\pA, c^{f}_\pA,
                           r^{g}_\pB, j^{p}_\pB, j^{r}_\pB, j^{v}_\pB, c^{f}_\pB}_i$
as input and predicts root trajectory velocities $ x^{traj}_i = \bracks{t^{\delta}_\pA, t^{\epsilon}_\pB}_i$. We train the regressor using reconstruction loss on both absolute root positions and root velocities. 
After training, we apply this network to the generated motions and replace the original root trajectories with the predictions based on the generated local motions.

\section{Experiments} \label{experiments}
We elaborate on our experimental setup, including dataset, baselines, ablations, and evaluation metrics. We provide the implementation details of our network components in the appendix.

\subsection{Dataset and Baselines}
\label{baseline}
We train and evaluate our model on the \datasetlong~\cite{siyao2023duolando}. It is currently the most comprehensive dataset of 3D two-person dance, comprising 10 distinct dance genres featuring intricate interactions between the dancers and approximately 1.9 hours of two-person dance motion data in the SMPLX representation~\cite{SMPL-X:2019} with corresponding music. The train-test split is $168{,}176$ frames and $42{,}496$ frames for two-person motions.

\paragraph{Data Pre-Processing}
As a pre-processing step, we augment the original dataset by generating sub-sequences using a sliding window of $400$ frames with a stride of $100$ frames. We sample the accompanying music at $15{,}360$ Hz to match the temporal dimension of the original motion sequence.
We further augment the dataset by interchanging persons $\pA$ and $\pB$ to create mirrored samples.
This results in a grand total of $4{,}556$ training samples and $1{,}144$ test samples, with a frame rate of $30$ FPS.
Finally, we process each two-person motion sequence by transforming the positions and orientations of $\pA$ and $\pB$ such that $\pA$'s root joint is at the global origin in the first frame and $\pA$'s body faces the $Z_+$ direction. We $Z$-normalize all the motion features before feeding them to the networks.

\noindent \paragraph{Baseline Setup}
We select the most relevant motion synthesis methods adaptable to two-person dance generation: Duolando~\cite{siyao2023duolando}, GCD~\cite{le2023controllable}, InterGen~\cite{liang2024intergen}, and MoFusion~\cite{dabral2022mofusion}.
GCD was originally trained on the AIOZ-GDANCE dataset~\cite{le2023music} for variable-size group choreography. We re-train it on \dataset~ for two-person dance generation with a thorough hyperparameter
search and report the best performances.
Duolando was initially designed to generate follower dance motions based on music signals and leader dance motions. To adapt it to two-person dance generation, we train a separate transformer model that generates the leader’s dance sequence from the input music. We then use these generated leader’s motions to condition the follower’s motions using the Duolando framework with pre-trained weights on \dataset~from the original work. InterGen, a diffusion-based approach,
originally generates two-person motions from text prompts.
To adapt its cooperative denoisers and mutual attention mechanisms
for music-driven dance generation, we retrain it on \dataset, replacing the original CLIP encoding~\cite{radford2021learning} with our music encoder network. 
For MoFusion,  we represent the two-person motion by concatenating the root-relative 3D joint position of each person. The second person is represented as a copy of the first person. The model is retrained on the \datasetlong.
Details on the preparation and training of these baselines are provided in the appendix.

\subsection{Ablated Versions} \label{sec:ablations}
We compare our proposed \method~ with the following ablations.
\begin{itemize}[leftmargin=*]
    \bulletitem \textbf{A1: w/o relational global positioning.} Represents root positions of both persons' motions in the global coordinates, following conventional approaches~\cite{liang2024intergen}.
    
    \bulletitem \textbf{A2: w/o hierarchical tokenization.} Tokenizes two-person dance motions using conventional single-layer VQ-VAE~\cite{zhang2023generating} and generates tokens through a single masked transformer.
    
    \bulletitem \textbf{A3: w/o unified representation.} Tokenizes each person's motion separately~\cite{javed2024intermask}, resulting in separate token sequences per person.
    
    \bulletitem \textbf{A4: w/o trajectory refinement module.} Removes the trajectory refinement module from our framework.
    
    \bulletitem \textbf{A5: alternate 438-D music encoding.} Uses 438-dimensional music representation~\cite{siyao2022bailando, siyao2023duolando} instead of MFCC and Chroma features.
    
    \bulletitem \textbf{A6: w/o music feature $z_m$ in VQ.} Trains the two-person VQ-VAE without conditional music signal encoding.
    
    \bulletitem \textbf{A7: alternate residual VQ.} Tokenizes the two-person dance sequences using four-layer Residual Quantization~\cite{guo2024momask}, with a dropout ratio of $0.2$.
    
    \bulletitem \textbf{A8: three-level token hierarchy.} Implements three hierarchies in the VQ-VAE.
 
\end{itemize}

\subsection{Evaluation Metrics}
\label{metrics}
We evaluate the two-person tokenizer's reconstruction quality using mean per joint position error (MPJPE) and mean per frame per joint velocity error (MPJVE)~\cite{ghosh2024remos} on the reconstructed motions after token decoding, along with the mean relative distance error (RDE) between the absolute root positions of $\pA$ and $\pB$.
On dance generation, we evaluate models across three aspects: individual motion quality, synchronization between dancers, and music-dance alignment.
For \textit{individual motion quality}, we compute the averaged Fr\'echet Inception Distance (FID)~\cite{heusel2017gans} over two persons, which measures the distributional difference between generated and ground-truth dance movement features. We also calculate latent variance to assess the diversity (Div) of generated dance movements, and
the physical foot contact score (PFC)~\cite{tseng2022edge} to measure the physical plausibility of the foot movements w.r.t. the ground plane.
For \textit{synchronization between dancers}, we use the PFID score~\cite{ng2022learning}, which compares the distributional difference between generated and ground-truth paired-motion features, instead of individual motion features.
We calculate contact frequency percentage (CF), defined as the proportion of frames where dancers are in contact with each other. Contact is determined by a minimum absolute distance of less than 40 cm between any joints of the two dancers.
For \textit{music-dance alignment}, we use the Beat Alignment Score (BAS) following prior works~\cite{siyao2022bailando, siyao2023duolando, bhattacharya2024danceanyway}, and report the averaged score among two persons.

\section{Results and Discussions}
\label{sec:result}
We summarize the results of our experiments and key observations.

\subsection{Quantitative Evaluation}
\label{subsec:quantitative_eval}
We evaluate our method quantitatively under two experimental settings: VQ reconstruction and two-person dance generation.

\paragraph{VQ Reconstruction}
\Cref{tab:reconstructions} reports the reconstruction quality of our hierarchical two-person motion VQ-VAE against the ablations and alternative designs discussed in \Cref{sec:ablations}. The vanilla VQ-VAE (ablation A2) produces noisy individual motion reconstructions (65.09 mm MPJPE) and struggles with high-fidelity close interactions (7.12 mm RDE). While applying 4-layer residual quantization in the middle of VQ-VAE (A7) partially mitigates these issues, the reconstruction quality remains suboptimal. In contrast, our hierarchical motion VQ-VAE, using only two quantization layers, demonstrates superior performance with high-fidelity reconstruction of both individual motions (36.26 mm MPJPE for person B) and interpersonal interactions (0.17 mm RDE).
Our motion representation designs contribute significantly to this improvement. Specifically, incorporating \textit{global relational positioning} and tokenizing \textit{unified two-person motions} reduces individual motion per-joint reconstruction errors by over 30\% and decreases the relative distance error from 7.12 to 0.17 millimeters. Further, our results demonstrate that incorporating conditional information (\textit{e.g.}, music) in the VQ-VAE enhances its reconstruction performance. While we explored adding more hierarchical levels, introducing a third level of tokens (A8) yielded only marginal gains, with some metrics showing slight degradation, likely due to increased learning complexity.

\paragraph{Two-Person Dance Generation}
\Cref{tab:quant_baseline} reports the performances of our method, the baselines, and the ablated versions on the \datasetlong. The baselines are retrained on the \datasetlong~using publicly available codes for these models.
\method~ demonstrates superior performance across key metrics, achieving the lowest individual FID score (1.31 versus GCD's 9.71) and paired-FID score (2.54 versus GCD's 8.11). Our method also achieves the highest music-dance alignment with a BAS score of 0.215, while maintaining reasonable contact frequency on par with ground-truth dances.
Ablation studies confirm the importance of both our hierarchical modeling (vs. ablation A2) and synchronization-dedicated representations (vs. A1 and A3). Removing either component leads to significant performance degradation, particularly in motion realism as measured by FID and PFID metrics. 
The two-hierarchy design provides the optimal balance between representational capacity and learning stability for duet dance generation, as opposed to training three-staged masked transformer (A8), where generation performance declines across key metrics.
Finally, trajectory refinement (vs. A4) provides modest overall improvements, notably enhancing foot contact plausibility. 

\begin{table}[t]
    \centering
    \caption{\textbf{Quantitative Evaluation of VQ Reconstruction}. Comparison of motion reconstruction quality after tokenization across ablated versions. 
    \textbf{Bold} indicates the best performance. 
    }
  
    \label{tab:reconstructions}
    \resizebox{\columnwidth}{!}{%
        \begin{tabular}{lccccc}
        \toprule
        Method & \multicolumn{2}{c}{MPJPE (mm) $\downarrow$}  & \multicolumn{2}{c}{MPJVE (mm) $\downarrow$} & RDE (mm) $\downarrow$ \\
        \cmidrule{2-6}
        & Person $\pA$ &  Person $\pB$ &  Person $\pA$ & Person $\pB$ & \\
        \midrule
        A1: w/o relational positioning & $59.35$ & $60.69$ & $14.33$ & $15.84$ & $7.23$ \\
        A2: w/o hier tokenization & $58.35$ & $65.09$ & $15.38$ & $17.54$ & $7.12$ \\
        A3: w/o unified representation  & $58.99$ & $59.33$ & $13.23$ & $13.34$ & $6.33$\\
        A6: w/o music feature $z_m$ & $39.60$ & $43.71$ & $12.82 $ & $14.68$ & $1.32$\\
        A7: Residual VQ (4 layers) & $49.77$ & $57.62$ & $12.51 $ & $14.49$ & $4.54$\\
        A8: $3$ level of hierarchy & $\mathbf{36.25}$ & $36.44$ & $\mathbf{10.81}$ & $12.05$ & $0.19$ \\
        \midrule
        Hier Two-Person VQ-VAE~(ours) & $ 36.32$ & $\mathbf{36.26}$ & $10.86$ & $\mathbf{11.84}$ & $\mathbf{0.17}$\\
        \bottomrule
        \end{tabular}
    }
\end{table}

\begin{table}[t]
    \centering
    \caption{\textbf{Quantitative Evaluation of Dance Generation.} Comparison of motion generation quality between baselines, ablated versions, and our method on the \datasetlong. \textbf{Bold} indicates best. 
    }
    \label{tab:quant_baseline}
    \resizebox{\columnwidth}{!}{%
        \begin{tabular}{lcccccc}
        \toprule
         Method & FID $\downarrow$ & Div $\rightarrow$ & PFID $\downarrow$ & PFC $\downarrow$ & CF $(\%) \rightarrow$ & BAS $\uparrow$ \\
        \midrule
        GT & $-$ & $15.67$ & $-$ & $0.36$ & $82.6$ & $0.213$ \\
        Duolando~\cite{siyao2023duolando} & $12.21$ & $14.72$ & $13.18$ & $16.22$ & $74.7$ & $0.202$ \\
        GCD~\cite{le2023controllable} & $9.71$ & $15.03$ & $12.03$ & $8.11$ & $78.1$ & $0.203$ \\
        InterGen~\cite{liang2024intergen} & $13.77$ & $15.01$ & $14.11$ & $12.4$ & $60.1$ & $0.172$ \\
        MoFusion~\cite{dabral2022mofusion} & $21.2$ & $15.60$ & $23.09$ & $7.5$ & $21.1$ & $0.202$ \\
        \midrule
        A1: w/o relational positioning & $5.03$ & $14.34$ & $14.97$ & $4.83$ & $79.5$ & $0.203$ \\
        A2: w/o hier tokenization & $4.77$ & $14.45$ & $15.44$ & $5.88$ & $80.2$ & $0.197$ \\
        A3: w/o unified representation & $5.65$ & $14.02$ & $18.99$ & $5.66$ & $75.66$ & $0.204$\\
        A4: w/o traj refinement & $2.62$ & $14.11$ & $2.81$ & $5.31$ & $78.2$ & $0.211$\\
        A5: $438$-D music representation & $2.45$ & $14.99$ & $2.87$ & $3.45$ & $72.1$ & $0.193$\\
        A8: 3 level of hierarchy & $1.55$ & $15.62$ & $2.95$ & $5.45$ & $70.1$ & $0.210$ \\
        \midrule
        \method~(ours) & $\mathbf{1.31}$ & $\mathbf{15.71}$ & $\mathbf{2.54}$ & $\mathbf{1.47}$ & $\mathbf{83.2}$ & $\mathbf{0.215}$ \\
        \bottomrule
        \end{tabular}
    }
\end{table}

\subsection{User Study}
\label{subsec:userstudy}
\begin{figure*}[t]
    \centering
    \includegraphics[width=0.6\linewidth]{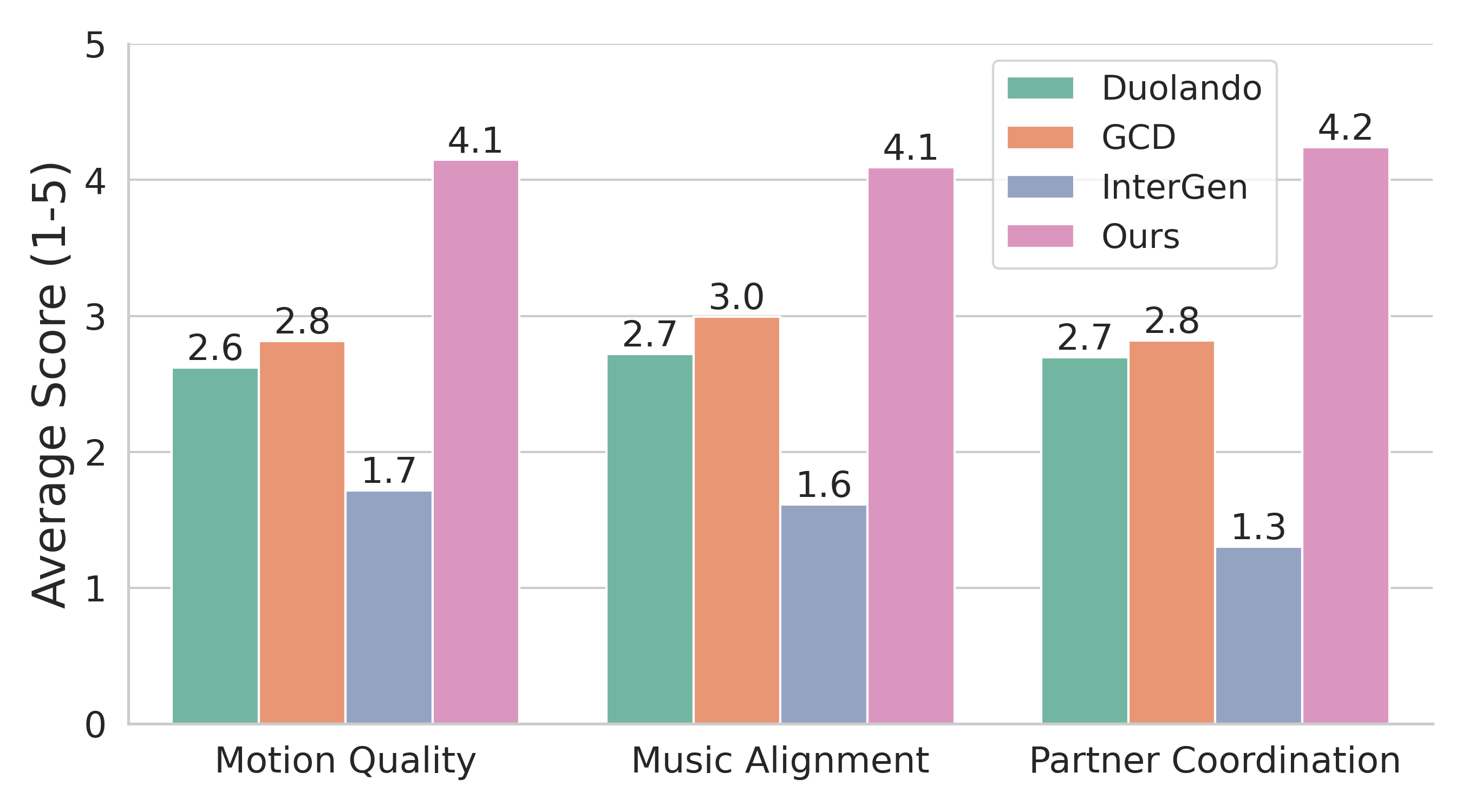}
    \caption{\textbf{User Study Results.} Each column indicates the average user rating on a 1-5 scale. \method~ consistently outperforms all baselines.}
    \label{fig:userstudy_results}
\end{figure*}

We conducted a user study to perceptually evaluate our method against the three baselines. We presented each participant with 10 sets of anonymized animations. In each set, we showed two-person dance animations generated by the different methods, all from the same music, in randomized orders. We showed all the animations side-by-side, and asked participants to rate each animation on \textit{motion quality}, \textit{music-motion alignment}, and \textit{partner coordination}, on 5-point Likert scales (more details in the appendix).
$30$ participants from different age groups and genders, with different levels of expertise in dancing, took our study. \Cref{fig:userstudy_results} summarizes the study results where we see \method~ earning the highest ranking across all three evaluation aspects, consistently achieving scores above `4' (at least $22\%$ higher than any other methods). At the other end, InterGen shows notably poor performance across all metrics (scoring below `2'), mainly due to its unnatural motion dynamics.

\subsection{Qualitative Results}
\begin{figure*}[t]
    \includegraphics[width=1.0\textwidth]{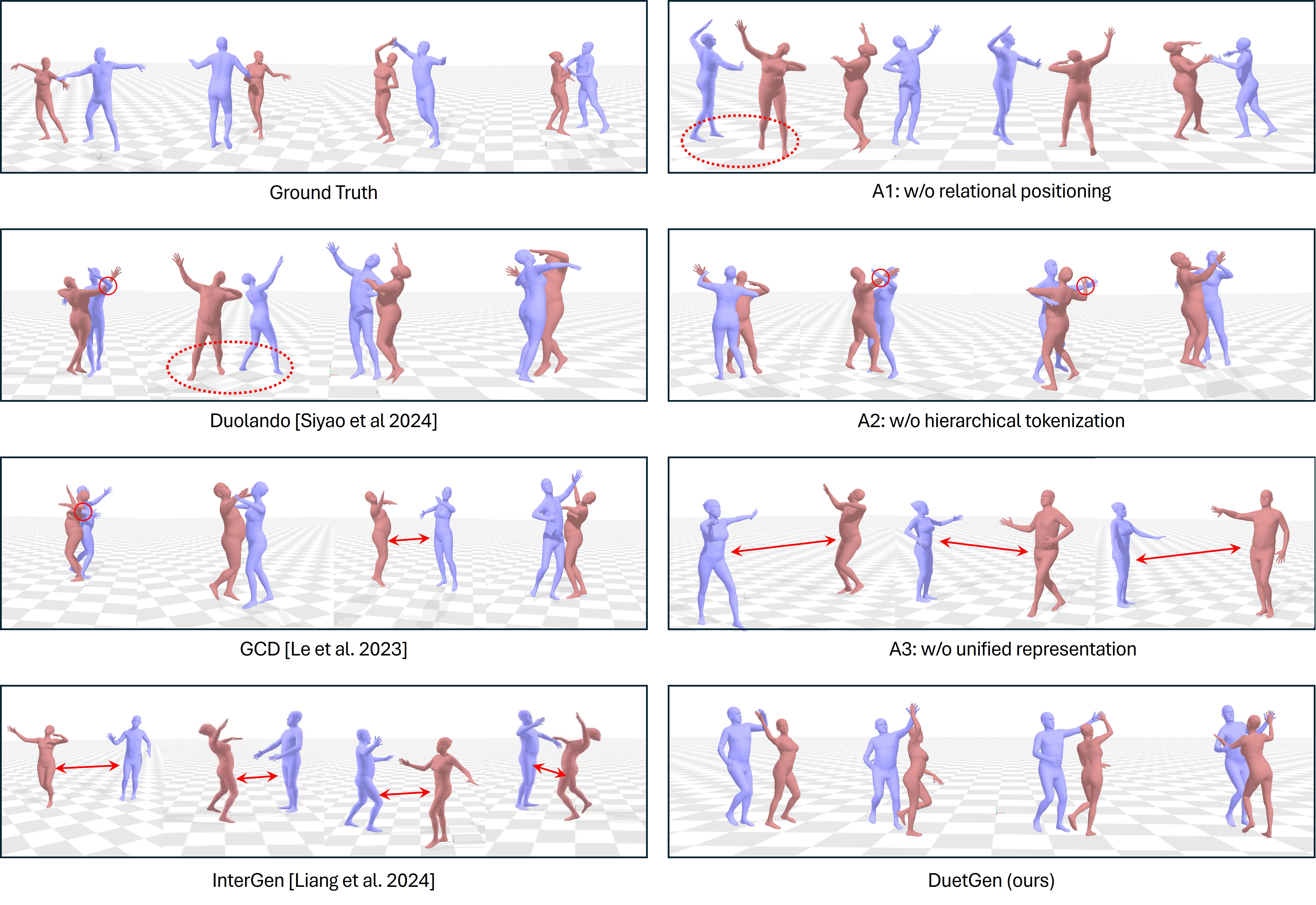}
    \caption{\textbf{Qualitative Comparisons.} Dance motions generated by \method, the baselines, and relevant ablations, from the same music input.
    Notice that the baseline methods and the ablations exhibit uncoordinated movements (\textit{red dots}), interpenetration (\textit{red circles}), and drift in root joint positions (\textit{red arrows}). In contrast, \method~ maintains natural interactions and well-synchronized two-person dance movements.
    }
    \label{fig:comparison}
\end{figure*}

\begin{figure*}[t]
    \includegraphics[width=1.0\textwidth]{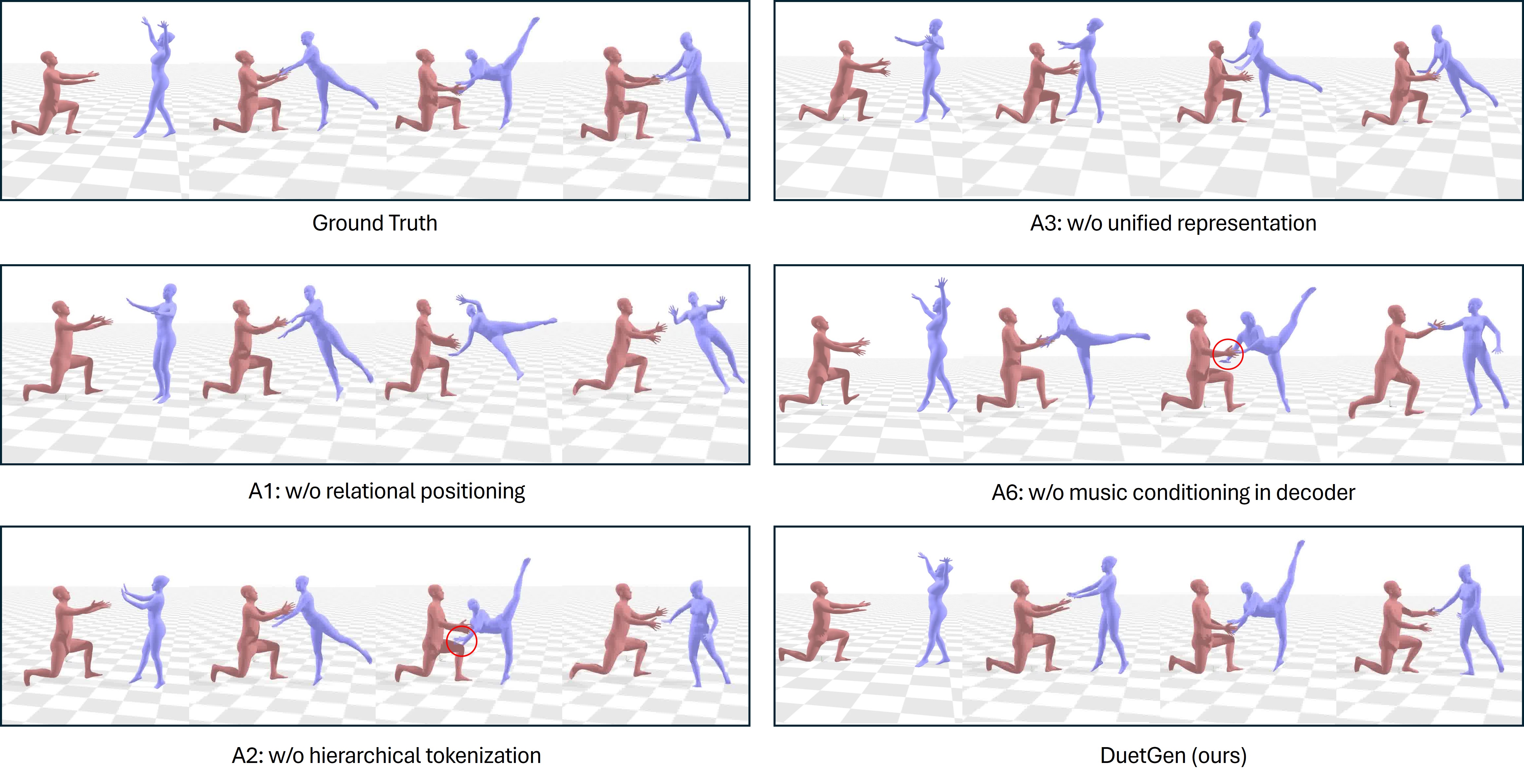}
    \caption{\textbf{Qualitative Comparison on VQ-VAE reconstruction.} Reconstruction quality of the hierarchical two-person VQ-VAE module of \method~compared to its ablations.
    Notice that the ablations exhibit uncoordinated movements and interpenetration (\textit{red circles}), while \method~ achieves interactions and synchronization between the two persons.}
    \label{fig:rec_comparison}
\end{figure*}

\Cref{fig:comparison} presents a qualitative comparison of dances generated by our method, the baselines, and the ablations.
InterGen~\cite{liang2024intergen} exhibits significant spatial drift, with dancers gradually moving far apart.
Duolando~\cite{siyao2023duolando} and GCD~\cite{le2023controllable} show improved partner awareness, but still suffer from poor coordination and frequent interpenetration between dancers.
Our ablations show that \textit{w/o relational positioning} (A1) and \textit{w/o unified representation} (A3), the dancers are mostly unaware of each other's motions, while \textit{w/o hierarchical tokenization} (A2) fails to model the fine-grained end-effector movements in close interactions. 
In contrast, \method~ produces well-coordinated dances, with natural synchronization and spatial coordination between the dancers throughout the sequence. 
Notably, when compared to ground-truth motions from the dataset, \method~ demonstrates the ability to generate novel dance sequences rather than merely reproducing the learning data.
We also compare the reconstruction quality of our hierarchical VQ-VAE with the ablations in \Cref{fig:rec_comparison}.

\section{Conclusion}
In this work, we present \method, the first approach for music-driven, two-person dance motion synthesis. Our approach comprises two principal components: (1) a hierarchical VQ-VAE to encode two-person dance motions into discrete tokens at two levels of granularity, capturing both global motion semantics and fine-grained articulations, and (2) the two-stage transformers to generate these two-level tokens from music using generative masked modeling. Our method produces two-person dance motions with plausible interactions and precise music-motion alignment, outperforming baselines in both quantitative evaluations and user studies.

\begin{figure*}[h]
    \centering
    \includegraphics[width=0.8\linewidth]{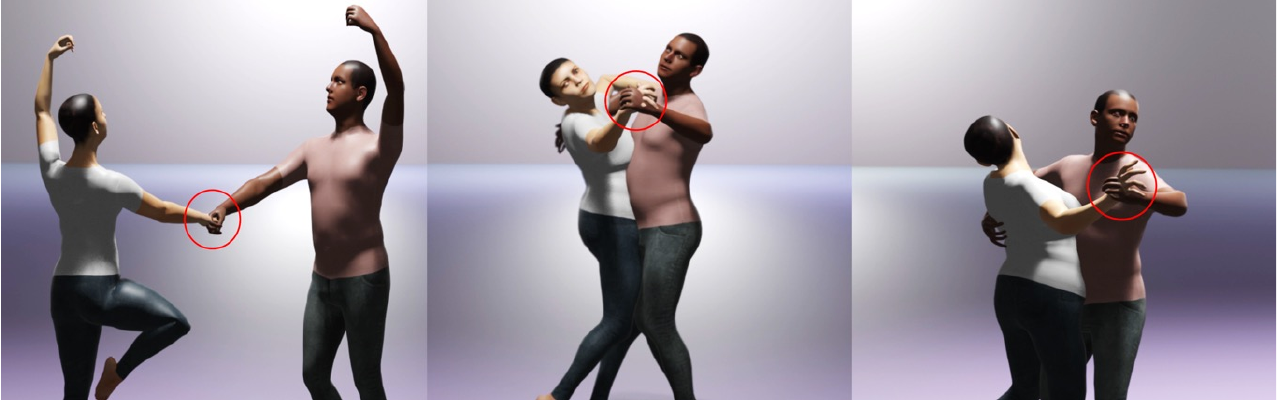}
    \caption{\textbf{Noisy Finger Motions in the \datasetlong.} Common artifacts in the \datasetlong~ include twisted or inter-penetrated finger motions (\textit{red circles}).}
    \label{fig:hand_noisy}
\end{figure*}

\paragraph{Limitations and Future Work}
While our method demonstrates significant advancements, it has certain limitations.
Our model does not explicitly address body shape variations, instead relying on ground-truth body shapes provided in the dataset. 
We currently do not model fine-grained finger interactions, primarily due to excessive noise in the captured finger motions in the \datasetlong~ (\Cref{fig:hand_noisy}).
Our model validation is also limited to the 1.9 hours of dance motion available in \dataset, currently the only available duet dataset with diverse genres of dance styles and music.
While collecting large-scale, high-quality duet dance datasets may be expensive, future works could explore learning from a combination of motion capture data and online dance videos to address the data scarcity.

\begin{acks}
This research was supported by Snap Inc., the EU Horizon 2020 grant Carousel+ (101017779), and the BMBF grant MOMENTUM (01IW22001).
\end{acks}

\bibliographystyle{ACM-Reference-Format}
\bibliography{main}
\newpage
\clearpage

\end{document}